\documentclass[12pt]{article}
\usepackage{latexsym,amsmath,amssymb,amsbsy,graphicx}
\usepackage[cp1251]{inputenc}
\usepackage[T2A]{fontenc}
\usepackage[space]{cite}

\makeatletter\def\@biblabel#1{\hfill#1.}\makeatother
\allowdisplaybreaks
\begin {document}

\noindent\begin{minipage}{\textwidth}
\begin{center}

{\Large{Simple finite-dimensional model of the metastable state}}\\[9pt]

{\large A.\,I. Dubikovsky$^{1a}$, P.\,K. Silaev$^{2b}$}\\[8pt]

\parbox{.96\textwidth}{\centering\small\it
{$^1$Department of Quantum Statistics and Field Theory,
Faculty of Physics, M.V.Lomonosov Moscow State University, Moscow 119991, Russia.}\\
\textit {$^2$Department of Quantum Theory and High Energy Physics,
Faculty of Physics, M.V.Lomonosov Moscow State University, Moscow 119991, Russia.}\\
\textit {E-mail: $^a$dubikovs@physics.msu.ru,
$^b$silaev314@yandex.ru}}\\[1cc]
\end{center}

\vspace{2pt}

{\parindent5mm 
We have constructed an approximate analytical solution of the spectral problem for a finite-dimensional matrix of a special kind, which turns out to be a very simple and quite satisfactory model of the metastable state. Most of the characteristic properties of the metastable state are reproduced: line shape, decay dynamics, and density of states. The correctness of the approximate analytical solution was verified by direct numerical calculations. The proposed model is a finite-dimensional analog of the Fano formalism.
\vspace{5pt}\par}


\vspace{5pt}
\end{minipage}

\section*{Introduction}
\mbox{}\vspace{-\baselineskip}

The model constructed in this work is a simple finite-dimensional analog of the Fano formalism \cite{fano0}, \cite{fano}. In the Fano formalism, the problem considered involves a discrete spectrum level corresponding to the unperturbed Hamiltonian \( H_0 \), which is ``embedded'' in the continuous spectrum.
Next, the perturbation Hamiltonian \( H_I \) is introduced, due to which the original discrete level becomes metastable, i.e., it transforms into a linear combination of generalized eigenvectors of the Hamiltonian \( H_0 + H_I \).
Here, the necessity inevitably arises to solve integral equations, with the solutions containing singular constructs such as \(\delta\)-functions and expressions of the form \(1/x\), understood in the sense of the principal value.
As a result, it becomes possible to obtain, first, the line shape for the metastable level; second, the density of states in the vicinity of the metastable level; and third, the exponential decay law of this metastable level.
However, it should be noted here that in some cases, the decay law may deviate from exponential (see \cite{fonda}).
Initially, this formalism was used to solve problems of autoionization, but later it proved useful in solving other problems as well.
In particular, this formalism turned out to be useful in describing the capture of slow neutrons by nuclei \cite{feshbach1}. However, for describing multichannel processes (``multichannel resonances''), it had to be somewhat generalized \cite{feshbach2}. 

Since the formalism itself is quite universal, it finds applications in describing a wide variety of physical systems.
It allows for the explanation of scattering, absorption, transmission spectra, and similar phenomena.
In particular, it has been used in studying polarization effects for photoelectrons \cite{kabachnik}, in investigating superconductors \cite{limonov1}, \cite{limonov2}, and semiconductors \cite{hopfield}, \cite{cerdeira}, as well as magnetic effects \cite{madhavan}.

The Fano formalism is widely used in describing light scattering by photonic structures.
In the work \cite{fan}, a narrow asymmetric resonance in waveguides with defects was studied.
In \cite{fan2}, the theory of Fano resonances in optical cavities is considered,
while \cite{kong} examines resonances in scattering by dielectric spheres with high permittivity.
It has been shown that Fano resonances are also observed in scattering by a single photonic crystal resonator \cite{galli}.
The Fano formalism is also applied in describing the effect of transparency induced by coupled resonators \cite{smith},
\cite{verslegers},
\cite{yang},
\cite{peng}.
Additionally, it is used to describe charge transfer processes through a quantum dot \cite{gores}, \cite{johnson}.
It is also employed in problems of quantum electrodynamics in the presence of strong fields \cite{grenier1}, \cite{grenier2}, \cite{kosta1}, \cite{kosta2}.

Strictly speaking, the Fano formalism is not the only way to obtain the line shape, density of states, and the decay law of a metastable level.
For example, one can use Fermi's golden rule (see, e.g., \cite{fermi}) to calculate the decay rate of a metastable state. After that, the line shape can be easily derived from the fact that the decay law and the line shape are related by a Fourier transform. However, this approach is not entirely consistent since, strictly speaking, Fermi's golden rule is only applicable for times \( t \ll \tau \), where \( \tau \) is the lifetime of the metastable state. 

An alternative and much more rigorous approach is based on the formalism of Green's functions.
An excellent exposition of this formalism can be found, for example, in \cite{khrust}.
This formalism allows one to either construct a closed system of equations for the matrix elements of the Green's function or, within perturbation theory, derive explicit expressions for these matrix elements.
In a system with an unperturbed Hamiltonian \( H_0 \), discrete levels correspond to the poles of the Green's function lying on the real axis. It can be shown that if a discrete level is embedded in a continuous spectrum, accounting for the perturbation \( H_I \) may shift the corresponding pole from the real axis downward --- that is, along the imaginary axis.
As a result, the corresponding state ceases to be an eigenvector of the Hamiltonian and instead becomes a metastable level --- a linear combination of the generalized eigenvectors of the Hamiltonian \( H_0 + H_I \).
Since the matrix elements of the evolution operator are expressed in terms of the matrix elements of the Green's function, it becomes possible to derive the decay law of this metastable level --- and not just for times \( t \ll \tau \).
Moreover, within the Green's function formalism, the line shape can be obtained not indirectly (via the decay law) but through direct computation. 

However, the approaches described above (the Fano formalism, the Green's function formalism) involve rather cumbersome calculations, which are typically carried out at a physical level of rigor.
In this work,
we construct a very simple model of a metastable level based on an approximate analytical solution of the spectral problem for a finite-sized Hermitian matrix.

The structure of the work is as follows: in Section 2 we construct an approximate analytical solution of the spectral problem, in Section 3 we obtain the line shape for the constructed model of a metastable level, in Section 4 we compare the approximate solution with the ``exact'' numerical solution for the spectral problem, in Section 5 we discuss the decay law of the metastable level in our model.

\section{Construction of the solution}
\mbox{}\vspace{-\baselineskip}

Let us consider a diagonal \( H_0 \) of size \( N \times N \) with an even \( N \). The requirement of evenness is not fundamental; it is imposed solely for convenience. We will treat the first basis vector as the initial discrete level. Since the energy is defined up to a constant, we can, without loss of generality, set the corresponding eigenvalue to zero:
\[
(H_0)_{11} = 0.
\]
The remaining basis vectors will be treated as an imitation of the continuous spectrum. To transform the initial discrete level into a metastable one, the discrete level must lie within the interval of the continuous spectrum. Therefore, we set
\[
(H_0)_{nn} \equiv E_n^{(0)} = dE (n - N/2 - 1),
\]
where \( n = 2, \ldots, N \). Thus, the energy of the discrete level lies at the center of the spectrum: for \( n_0 = N/2 + 1 \), the energy \( E_{n_0} = 0 \).
For a satisfactory imitation of the continuous spectrum, two conditions must be met:
sufficiently small \( dE \), and
sufficiently large interval over which the eigenvalues (imitating the continuous spectrum) are distributed, i.e., a sufficiently large \( N \cdot dE \).
The second condition is necessary to ensure that this interval encompasses all vectors significantly affected by the interaction with the discrete level. Both conditions will be refined in the course of constructing the solution.

When we solve the problem for the perturbed Hamiltonian, we will number the \mbox{eigenvectors} and eigenvalues in order of increasing energy.
To facilitate comparison with the unperturbed spectrum, the latter must be ordered. In the ordered unperturbed spectrum:
the lowest level \( E^{(0)}_{min} = -dE (N/2 - 1) \) is the first,
\( E^{(0)}_{max} = dE (N/2 - 1) \) is the last,
while at the center of the ordered spectrum lies a doubly degenerate level with \( E^{(0)} = 0 \).

The perturbation \( H_I \) should describe transitions from level \( (i=1) \) to level ``\( f \)'' for \( f \geq 2 \).
Let us assume that the corresponding off-diagonal matrix element does not depend on the index ``\( f \)'':
\[
(H_I)_{1f} = (H_I)_{f1} = W
\]
for \( f = 2 \ldots N \). The constant \( W \) can always be made real by exploiting the freedom in choosing the phase of the basis vectors from the 2nd to the \( N \)-th.
The fact that the matrix elements are independent of ``\( f \)'' corresponds to the condition that the matrix element \( \langle \psi_f | H_I | \psi_i \rangle \) should not vary significantly for those states ``\( f \)'' of the continuous spectrum whose energies \( E_f \) lie within a small neighborhood of the initial state's energy \( E_i \). Since \( H_I \) is a perturbation, the constant \( W \) must be small.

Let us construct an approximate analytical solution for the Hamiltonian \( H = H_0 + H_I \) --- both for the eigenvectors and eigenvalues.

If the eigenvalue \( E_k \) is known, then the corresponding eigenvector can easily be found. Indeed, let us take the stationary Schr\"{o}dinger equation
\[
H|\psi^{(k)}\rangle = E_k |\psi^{(k)}\rangle
\]
and use the rows from the 2nd to the \( N \)-th.
We then obtain:
\begin{equation}
\psi_n^{(k)} = - \psi_1^{(k)} \cdot \frac{W}{E_n^{(0)} - E_k} \quad \text{for} \quad n=2\ldots N.
\label{ostalnye}
\end{equation}
Accordingly, \( \psi_1^{(k)} \) can be determined from the normalization condition:
\[
1 = \sum_{n=1}^N \left(\psi_n^{(k)}\right)^2 = \left(\psi_1^{(k)}\right)^2 \cdot \left( 1 + \sum_{n=2}^N \frac{W^2}{(E_n^{(0)} - E_k)^2} \right).
\]
This is an exact equality. However, the sum \( \sum_{n=2}^N \) can be approximately computed as:
\[
\sum_{n=2}^N \frac{W^2}{(E_n^{(0)} - E_k)^2} \approx \frac{W^2}{dE^2} \cdot \left( \frac{\pi^2}{\sin^2\left(\pi E_k / dE\right)} - \right. \hbox{\strut\kern 10 em}
\]
\begin{equation}
\hbox{\strut\kern 4 em} \left. - \frac{1}{N/2 - 1/2 - E_k / dE} + \frac{1}{1/2 - N/2 - E_k / dE} \right).
\label{ocenkanormy}
\end{equation}
This approximate formula follows from the exact formula
\[
\sum_{n=-\infty}^{+\infty} \frac{1}{(n - a)^2} = \frac{\pi^2}{\sin^2(\pi a)}.
\]
The difference between this infinite sum and the sum from ``2'' to ``\( N \)'' consists of two ``semi-infinite'' sums (from \(-\infty\) to 1 and from \(N+1\) to \(+\infty\)). Since \( dE \) is small, they can be treated as integral sums. Moreover, it is appropriate to use a second-order accuracy formula, where the function value is taken at the midpoint of each interval \( dE \). Thus, both semi-infinite sums are estimated using integrals, which yield the second and third terms in the parentheses in relation (\ref{ocenkanormy}).

It might seem that replacing the integral sum with an integral for a function of the form \(1/(x-a)^2\) would lead to significant errors when the singularity at ``\(a\)'' is located close to the integration region. In other words, one might expect a large deviation of the approximate result (\ref{ocenkanormy}) from the exact sum for eigenvalues lying near the ends of the interval, i.e., for \(k \sim 1\) and \(k \sim N\).
This reasoning is generally correct. However, as we will see later, the value of \(E_k\) does not differ too much from \(E_k^{(0)}\). Moreover, at the edges of the energy interval, this difference is minimal. This is to be expected, since the greater the difference between the energy level \(E_k^{(0)}\) and the discrete level energy \(E = 0\), the smaller the perturbation introduced by the perturbation Hamiltonian \(H_I\) into the energy \(E_k^{(0)}\).

As a result,
at the edges of the interval (for \(k \sim 1\) and \(k \sim N\)), the difference \(E_k - E_k^{(0)}\) is extremely small, and the dominant contribution to the sum comes from the term with \(n = k\). Accordingly, in the approximate expression (\ref{ocenkanormy}), the main contribution is given by the first term in the parentheses, while the second and third terms play no significant role.

Conversely, in the middle of the interval (\(k \sim N/2\)), the difference \(E_k - E_k^{(0)}\) is not too small, but precisely at the midpoint of the interval replacing the integral sum with an integral is valid.
Therefore, the approximate expression (\ref{ocenkanormy}) can be applied across the entire range of indices \(k\).

These arguments were verified by direct numerical calculations and were found to be correct.

Thus, for a given eigenvalue \( E_k \), the eigenvector is given by expression (\ref{ostalnye}), where

\[
\psi_1^{(k)} \approx \left\{1 + \frac{W^2}{dE^2} \cdot \left( 
\frac{\pi^2}{\sin^2\left(\pi E_k / dE\right)} - 
\frac{1}{N/2 - 1/2 - E_k / dE} + \right.\right. 
\]

\begin{equation}
\hbox{\strut\kern 2 em} \left.\left.
\phantom{\frac{\pi^2}{\sin^2\left(\pi E_k / dE\right)}} +
\frac{1}{1/2 - N/2 - E_k / dE}
\right) \right\}^{(-1/2)}.
\label{pribnorma}
\end{equation}

Now we need to find the energies \( E_k \).
Let us use the first line of the stationary Schr\"{o}dinger equation:
\[
E_k \cdot \psi_1^{(k)} = \sum_{n=1}^N H_{1n} \cdot \psi_n^{(k)} = \sum_{n=2}^N W \cdot \left( -\psi_1^{(k)} \frac{W}{(E_n^{(0)} - E_k)} \right).
\]
From this, we obtain the equation for \( E_k \):
\[
E_k = - \sum_{n=2}^N \frac{W^2}{(E_n^{(0)} - E_k)} \; .
\]
The sum on the right-hand side can be approximated as:
\[
- \sum_{n=2}^N \frac{W^2}{(E_n^{(0)} - E_k)} \approx \frac{W^2}{dE} \cdot \left\{ \pi \cot(\pi E_k / dE) - \log\left( \frac{N - k + 1/2}{k - 1/2} \right) \right\} .
\]
This estimate for the sum is obtained in exactly the same way as the previous one, except that one should take as the main formula the exact expression for the infinite sum:
\[
\sum_{n=-\infty}^{+\infty} \frac{1}{n - a} = \frac{\pi \cos(\pi a)}{\sin(\pi a)} \; .
\]
Thus, we need to solve the approximate equation:
\begin{equation}
E_k = \frac{W^2}{dE} \cdot \left\{ \pi \cot(\pi E_k / dE) - \log\left( \frac{N - k + 1/2}{k - 1/2} \right) \right\} .
\label{pribura}
\end{equation}
Let us define the following quantities:
\begin{align*}
& \vrule width 0mm depth 1.8ex \Gamma \equiv 2\pi W^2 / dE \, , \\ 
& \vrule width 0mm depth 1.8ex E_k^{(I)} = dE (k - N/2 - 1/2), \\
& \vrule width 0mm depth 3.4ex E_k^{(II)} = - \frac{dE}{\pi} \arctan\left( \frac{dE (k - N/2 - 1/2)}{\Gamma / 2} \right), \\
& \vrule width 0mm depth 3.8ex E_k^{(III)} = - \frac{dE}{\pi} \arctan\left( \frac{dE (k - N/2 - 1/2) + E_k^{(II)}}{\Gamma / 2} \right), \\
& E_k^{(IV)} = - dE \log\left( \frac{N - k + 1/2}{k - 1/2} \right) \cdot \frac{1}{\pi^2 + \left( (E_k^{(I)} + E_k^{(III)}) dE / W^2 \right)^2} \; . 
\end{align*}
We will solve equation (\ref{pribura}) using the method of successive approximations.
The zeroth-order approximation to the solution is:
\begin{equation}
E_k \approx E_k^{(I)} + E_k^{(II)}.
\label{nulevoeprib}
\end{equation}
In this approximation, we neglect the logarithmic term, and the left-hand side retains an uncompensated term proportional to \( E_k^{(II)} \).

Replacing \( E_k^{(II)} \) with \( E_k^{(III)} \) allows this term to be compensated.
Generally speaking, it might seem that for even greater self-consistency of the solution, one should replace \( E_k^{(II)} \) in the argument of the arctangent in the definition of \( E_k^{(III)} \) with \( E_k^{(III)} \) itself.
However, this would exceed the required accuracy. Numerical experiments show that such a replacement practically does not change the difference between the approximate analytical solution and the ``exact'' numerical one.

Finally, we need to find the correction associated with the logarithmic term. Assuming
$$ E_k \approx E_k^{(I)} + E_k^{(III)} + \delta E_k, $$
we find that \(\delta E_k = E_k^{(IV)}\).
Thus, the final expression is:
\begin{equation}
E_k \approx E_k^{(I)} + E_k^{(III)} + E_k^{(IV)} \; .
\label{pribresh}
\end{equation}
This approximate analytical solution was verified by direct numerical computation. It was found to be highly accurate.

It should be noted that we have now justified the statement about the difference between \( E_k \) and \( E_k^{(0)} \). For \( k \sim 1 \) and \( k \sim N \), the value of the ``\(\arctan\)'' function in \( E_k^{(d)} \) turns out to be of the order of \(\mp \pi/2\), so the addition \(\pm dE/2\) cancels out the ``\(-dE/2\)'' term present in \( E_k^{(I)} \).
As a result, we obtain \( E_1 \approx -dE(N/2 - 1) = E^{(0)}_{min} \) and \( E_N \approx dE(N/2 - 1) = E^{(0)}_{max} \). Thus, as expected, the perturbation \( H_I \) barely affects the energy of the highest and lowest levels. This is because their unperturbed energy is significantly different from the energy of the original discrete level with which they ``interact''.
Only the energy of levels close to zero --- i.e., those with indices \( k \sim N/2 \) --- undergoes noticeable changes. Specifically, around the index \( k = N/2 \), there exists a region of width approximately \( \Gamma/dE \). Within this region, the energy correction varies from roughly \( dE/2 \) to roughly \(-dE/2\).
This corresponds to an increase in the density of states near the metastable level, with the number of states increasing by one. Initially, the energy interval \([E^{(0)}_{min}, E^{(0)}_{max}]\) contained \((N-1)\) levels, with the \( E = 0 \) level being doubly degenerate. As a result of the perturbation, the same interval now contains \( N \) non-degenerate levels.

The ``compression'' of levels is described by the formula for the density of states at a given energy, \(\rho(E_k) = dk/dE_k\).
In this case,
\begin{equation}
\frac{1}{\rho(E_k)} = \frac{dE_k}{dk} \approx dE - \frac{dE^2}{\pi} \;\frac{\Gamma/2}{\left(E_k^{(I)} \right)^2 + \Gamma^2/4} \; .
\label{plotnostur}
\end{equation}
In this expression, we have used the zeroth-order approximation for the energy (\ref{nulevoeprib}). Here, it is important to emphasize that the quantity \( E_k^{(a)} \) in the denominator is not the absolute energy value but rather the deviation of the energy from the original discrete level's energy --- i.e., from zero.
This can be easily verified by repeating all the derivations while setting \( E^{(0)}_1 = \mathcal{E}_0 \) and
\( E_n^{(0)} = \mathcal{E}_0 + dE(n - N/2 - 1) \).

It can be seen that the reduction in the interval between neighboring levels is described by a Lorentzian curve. This is a characteristic property of all metastable levels.
For instance, metastable levels associated with scattering resonances (``quasi-normal modes'') exhibit the same property. Indeed, in the presence of a resonance, the solution to the radial Schr\"{o}dinger equation must have the asymptotic form:
$$
\psi_\ell(r \to \infty) = D \cos\left(pr + \delta_\ell^{(0)} - \arctan\left(\frac{p_1}{p - p_0}\right)\right),
$$
where \(\delta_\ell^{(0)}\) is the background scattering phase, \(p_0\) determines the resonance position, and \(p_1\) its width.
Such behavior of the scattering phase arises when the pole of the scattering amplitude on the complex plane of the wave number \(p\) lies at the point \(p = p_0 - i p_1\).

Let us transition from the continuous spectrum to a discrete one by imposing the Dirichlet condition at \(r = L\) (where \(L \to \infty\)). In doing so, it should be noted that we are considering a small neighborhood around the resonance \(p = p_0\). Then, we find that the interval between levels far from the resonance is
$$
dE = \frac{\hbar^2 p_0}{m} \frac{\pi}{L},
$$
and \(dE_k/dk = \frac{\hbar^2 p_0}{m} dp_k/dk\).
Additionally, the relationship between \(p_1\) and \(\Gamma\) must be used:
$$
\frac{\hbar^2 k_0}{m} k_1 = \frac{\Gamma}{2}.
$$
Then, we finally obtain
$$
\frac{dE_k}{dk} = dE - \frac{dE^2}{\pi} \frac{\Gamma/2}{(E_k - E_0)^2 + \Gamma^2/4},
$$
which indeed exactly coincides with (\ref{plotnostur}).

\section{Shape of the line}
\mbox{}\vspace{-\baselineskip}

Let us use a very crude approximation in (\ref{pribnorma}):

$$
\psi_1^{(k)} \approx \left\{ \frac{W^2}{dE^2} \cdot \left( \frac{\pi^2}{\sin^2\left(\pi E_k / dE \right)} \right) \right\}^{(-1/2)}.
$$
Here, we have neglected the term ``1'' compared to the term containing the factor \( W^2/dE^2 \), and we have also discarded corrections related to the fact that, when computing the norm, the summation runs not from ``\(-\infty\)'' to ``\(+\infty\)'', but rather from ``\(2\)'' to ``\(N\)''.

Let us substitute the zeroth-order approximation for the energy \( E_k \approx E_k^{(I)} + E_k^{(II)} \) into this approximate expression. Then we obtain:
\begin{equation}
\left|\psi^{(k)}_1\right|^2 \approx \frac{W^2}{ \left(E_k^{(I)}\right)^2 + \Gamma^2/4 } = dE \cdot \frac{\Gamma }{2\pi} \frac{1}{ \left(E_k^{(I)}\right)^2 + \Gamma^2/4 } \; . \label{formalinii3}
\end{equation}
Let us emphasize once again that, in reality, the denominator does not contain the absolute value of the energy but rather:
$$
\left(E_k^{(I)}\right) = \left(E_k^{(I)} - 0\right) = \left(E_k^{(I)} - \mathcal{E}_0\right),
$$
i.e., the deviation of the energy \( E_k^{(I)} \) from the energy of the original discrete level \( \mathcal{E}_0 = 0 \).
This means that, to some approximation, our model reproduces the answer for the line shape.

Indeed, the quantity \(\left|\psi^{(k)}_1\right|^2\) represents the expansion coefficient of the original discrete level, which, due to the perturbation \(H_I\), has become metastable. This coefficient is expressed in terms of the eigenvectors of the Hamiltonian \(H_0 + H_I\), which simulate a continuous spectrum. Thus, this expression models the probability density of having energy \(E_k\), i.e., the line shape.
As expected for a metastable level, the line shape turns out to be a Lorentzian curve. Moreover, the quantity \(\Gamma\) that we introduced is consistent with the transition rate formula, i.e., Fermi's golden rule:
$$
\Gamma = 2\pi \sum_f |\langle \psi_f | H_I | \psi_i \rangle|^2 \delta(E_f - E_i) = 2\pi W^2 / dE,
$$
since, when transitioning from a discrete to a continuous spectrum, we must assume that \(\delta(0) = 1/dE\).

\section{Comparison of the approximate analytical solution with the numerical one}
\mbox{}\vspace{-\baselineskip}

The spectral problem for a finite-sized matrix can easily be solved numerically with very high accuracy.
This allows us to estimate the errors in our approximate analytical solution.
Numerical experiments show that these errors are of two types.

The first type of errors are boundary errors.
Both in the expression for energy (\ref{pribresh}) and in the expression for the norm (\ref{pribnorma}), they reach their maximum at the ends of the interval, i.e., for \(k \sim 1\) and \(k \sim N\).
Boundary errors arise when the value of \(N\) is not sufficiently large, meaning we are considering too short a segment of the ``continuous spectrum'' around the ``discrete level''. In other words, a situation occurs where the perturbation still has a noticeable effect on the minimum energy \(E^{(0)}_{min}\) and the maximum energy \(E^{(0)}_{max}\).
When constructing the approximate solution, we already noted that our approximate estimates for the sums are only applicable if, at the ends of the energy interval (i.e., for \(k \sim 1\) and \(k \sim N\)), the quantity \(\left| E_k - E^{(0)}_{k} \right|\) --- the difference between the perturbed and unperturbed spectra --- is sufficiently small.
The size of the region where the perturbation has a noticeable effect on the spectrum is characterized by the dimensionless parameter \(R \equiv \Gamma/dE \sim W^2/dE^2\).
For boundary errors to be small, the condition \(N \gg R\) must hold.
From this, it is clear that for a fixed \(R\), boundary errors decrease as \(N\) increases.
Conversely, for a fixed \(N\), boundary errors grow as \(R\) increases.

The second type of errors, the ``mid-interval'' errors, on the contrary, arise in the middle of the energy interval, i.e., for \(k \sim N/2\). They are localized within a region of size on the order of \(R = \Gamma/dE\).
These errors occur where the influence of the perturbation is strongest and are related to the fact that, in the course of calculations, we replaced summation with integration. Consequently, they should decrease as \(dE\) decreases, i.e., as the dimensionless parameter \(R\) increases. Numerical experiments confirm this conclusion.

For clarity, we have provided graphs showing the dependence of errors on the parameter \( R \) for three fixed values of \( N \): \( N = 2000 \), \( N = 4000 \), and \( N = 8000 \).
  
The error in the energy expression \(\Delta_1\) is defined as the maximum deviation, across all \(k\), of the approximate expression (\ref{pribresh}) from the corresponding ``exact'' numerical value.
 
Since the square of \(\psi^{(k)}_1\) rather than \(\psi^{(k)}_1\) itself has direct physical meaning, we define the error in the norm expression \(\Delta_2\) as the maximum deviation, across all \(k\), of the quantity \(\left|\psi^{(k)}_1\right|^2\) (where \(\psi^{(k)}_1\) is taken from (\ref{pribnorma})) from the corresponding numerical value.
 
Finally, the third error \(\Delta_3\) does not characterize the quality of our constructed approximate analytical solution but rather the quality of our model. We define \(\Delta_3\) as the maximum deviation, across all \(k\), of the Lorentzian curve (\ref{formalinii3}) from the exact numerical value for \(\left|\psi^{(k)}_1\right|^2\).
It is clear that the model can only be considered satisfactory if the Lorentzian line shape for the metastable level is reproduced with sufficient accuracy.

\begin{figure}
\centerline{\includegraphics[scale=0.7]{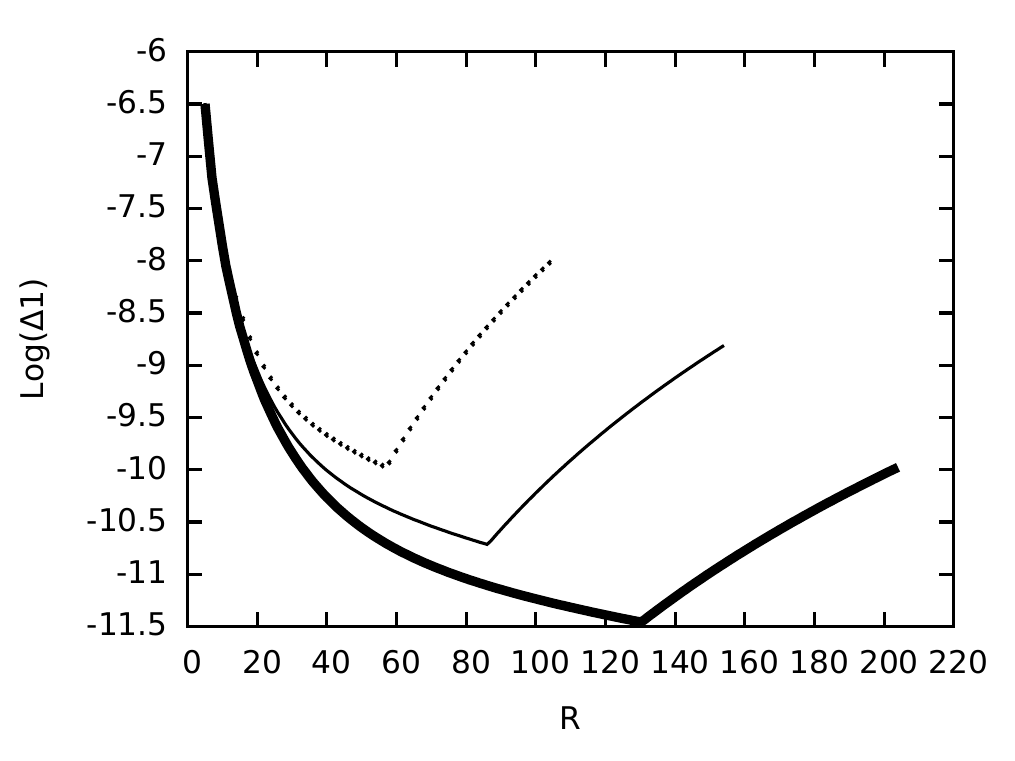}}
\caption{The logarithm of the error \(\Delta_1\) as a function of the parameter \(R\). The bold line corresponds to \(N=8000\), the thin line to \(N=4000\), and the dotted line to \(N=2000\).}
\label{err1}
\end{figure}

\begin{figure}
\centerline{\includegraphics[scale=0.7]{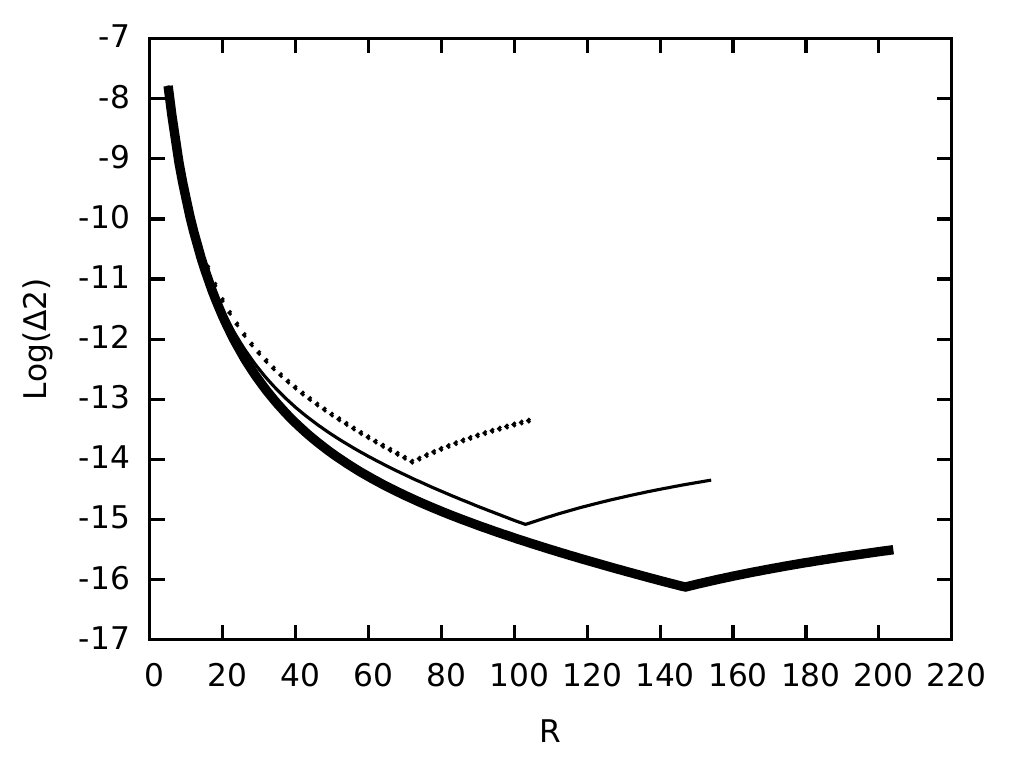}}
\caption{The logarithm of the error \(\Delta_2\) as a function of the parameter \(R\). The bold line corresponds to \(N=8000\), the thin line to \(N=4000\), and the dotted line to \(N=2000\).}
\label{err2}
\end{figure}

\begin{figure}
\centerline{\includegraphics[scale=0.7]{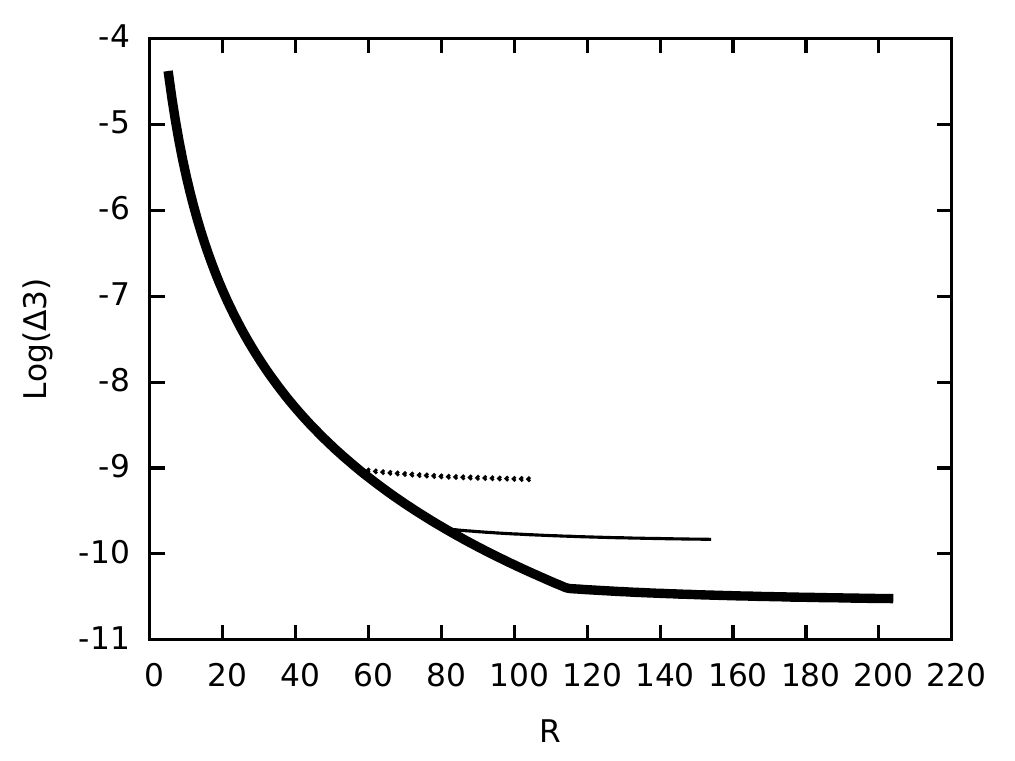}}
\caption{The logarithm of the error \(\Delta_3\) as a function of the parameter \(R\). The bold line corresponds to \(N=8000\), the thin line to \(N=4000\), and the dotted line to \(N=2000\).}
\label{err3}
\end{figure}
 
From Figures \ref{err1}---\ref{err3}, it can be seen that the dependence of the errors on the parameter \( R \) for a fixed \( N \) is practically uniform. Initially, i.e., for relatively small \( R \), all three errors decrease as \( R \) increases. In this case, the errors at the edges of the energy interval (i.e., for \( k \sim 1 \) and \( k \sim N \)) are much smaller than the errors in the middle of the interval (i.e., for \( k \sim N/2 \)).
However, as we have already mentioned, as \( R \) increases, it is the ``middle'' errors that decrease, while the edge errors grow. Therefore, at a certain value \( R = R_0 \), the edge errors become equal to the middle errors. With further increase in \( R \), the maximum error shifts to the edges of the interval and begins to rise.
Thus, the minimal error is achieved at a sort of ``turning point'' \( R = R_0 \). It should be noted that, as expected, the value of \( R_0 \) increases with \( N \). After all, the larger \( N \) is, the later edge effects may manifest.
In Figures \ref{err1} and \ref{err2}, the turning points and their dependence on \( N \) are clearly visible. As for the error \( \Delta_3 \) (see Fig. \ref{err3}), the turning points are also well-defined, but after each turning point, \( \Delta_3 \) does not start to increase --- it simply ceases to decrease noticeably.

As for the dependence of errors on \( N \) at a fixed \( R \), it is clearly seen that they initially decrease as \( N \) grows, but with further increase in \( N \), the errors stop decreasing --- i.e., a kind of saturation occurs. This was also to be expected, since at a fixed \( R \), the size of the region where the perturbation affects the spectrum is also fixed. As long as edge effects play any role, increasing \( N \) leads to a reduction in errors. However, further growth in \( N \) results in the addition of states that are entirely unaffected by the perturbation \( H_I \). In this case, the maximum errors remain ``middle'' ones and thus cease to depend on \( N \).

Since all three plots are necessarily presented on a logarithmic scale, we also provide in Table~\ref{tabl1} the absolute values of the errors at the turning points --- i.e., the minimum error values for each \( N \) --- for clarity.

\begin{table}[htbp]
\caption{Absolute error values at the turning points}
\begin{center}
\begin{tabular}{|c|c|c|c|c|c|c|}\hline
$N$ & $R_0^{(1)}$ & $\Delta_1$ &  $R_0^{(2)}$ & $\Delta_2$ & $R_0^{(3)}$ & $ \Delta_3$    \\ \hline 
2000  &  57.2   & $4.6\cdot10^{-5}$   
& 72.2 &  $8.0\cdot10^{-7}$  &  56 & $1.2\cdot10^{-4}$    \\ 
4000  &  86.2 & $2.2\cdot10^{-5}$     & 103.2 & $2.8\cdot10^{-7}$  & 81 & $6.1\cdot10^{-5}$   \\ 
8000 & 130.3 & $1.1\cdot10^{-5}$     & 147.0 & $9.9\cdot10^{-8}$  & 126  & $2.9\cdot10^{-5}$   \\ 
\hline
\end{tabular}\label{tabl1}
\end{center}
\end{table}

Thus, both the errors of the approximate solution itself, \(\Delta_1\) and \(\Delta_2\), and the error of the constructed model, \(\Delta_3\), are sufficiently small even for a moderate \(N = 2000\). Moreover, they decrease quite rapidly with a simultaneous increase in \(R\) and \(N\).

\section{Dynamics of the decay of a metastable level}
\mbox{}\vspace{-\baselineskip}

As for the decay dynamics of the state $|1\rangle$ that we constructed, which models a metastable level, the deviation from the usual exponential decay law is caused by the same factors that lead to the appearance of edge and midpoint errors in the approximate solution.

The finite-dimensional nature of the model leads, firstly, to the fact that the spectrum exists within a finite energy interval \( E^{(0)}_{\text{min}} < E < E^{(0)}_{\text{max}} \). As a result, at times \( t < T_{\text{min}} \sim \hbar / E^{(0)}_{\text{max}} \), a constant decay rate \( \Gamma / \hbar \) cannot be achieved.
Moreover, the decay law will exhibit small oscillations with a frequency \( \sim 2\pi \hbar / E^{(0)}_{\text{max}} \).
For a constant decay rate to emerge, the energy distribution must extend to infinity and decrease as \( \sim 1/E^2 \).

Secondly, the finite-dimensional nature of the model results in a discrete spectrum with a characteristic energy level spacing \( dE \).
Consequently, the dynamics of the finite-dimensional model can accurately simulate the dynamics of a system with a continuous spectrum only for times \( t \ll T_{\text{max}} \sim \hbar / dE \). In particular, the dynamics of the unperturbed system are strictly periodic with a period \( T_0 = 2\pi\hbar / dE \).

\begin{figure}
\centerline{\includegraphics[scale=0.7]{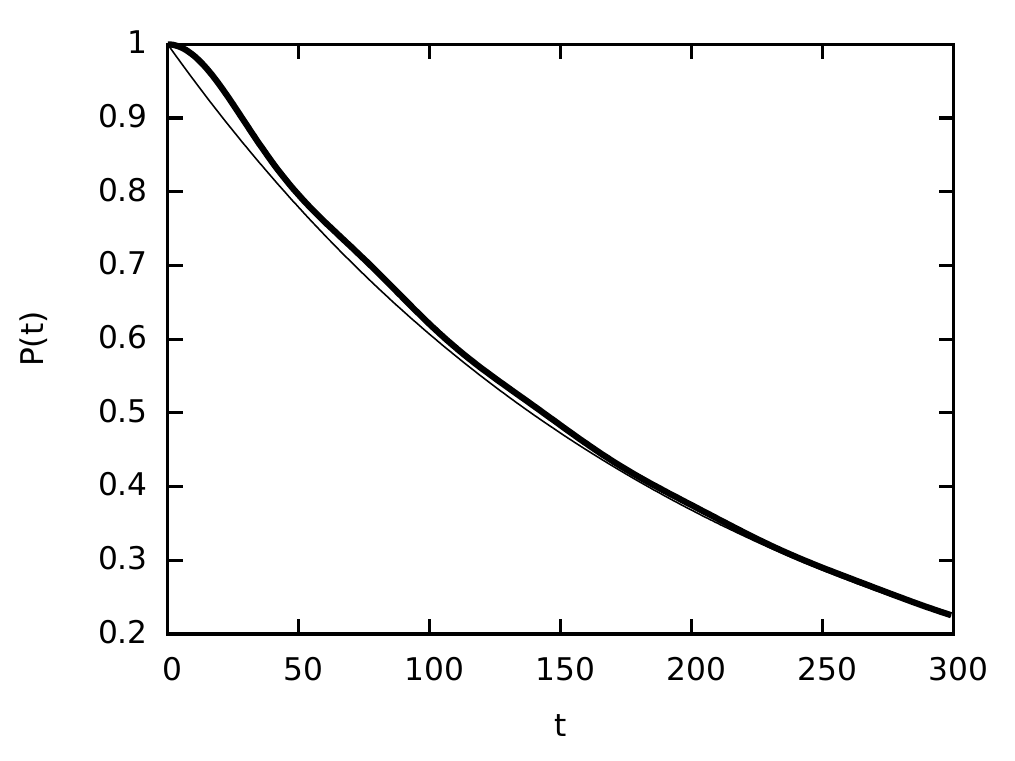}}
\caption{The decay law \(P(t)\) for the metastable level model \(|1\rangle\) as a function of time \(t\). The bold line represents \(P(t)\), while the thin line corresponds to the exponential \(\exp(-\Gamma t/\hbar)\).}
\label{risraspad1}
\end{figure}

\begin{figure}
\centerline{\includegraphics[scale=0.7]{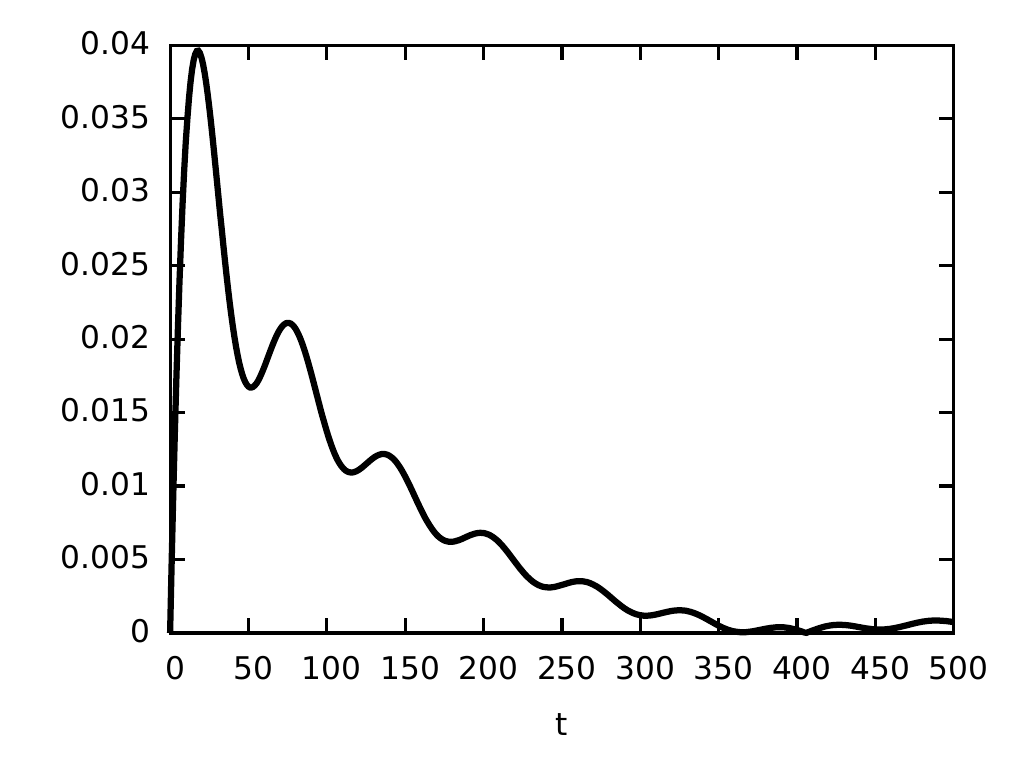}}
\caption{Deviation of the decay law \(P(t)\) from the exponential \(\exp(-\Gamma t/\hbar)\) as a function of time \(t\).}
\label{risraspad2}
\end{figure}

Let us derive the decay law for the metastable level model \(|1\rangle\). If \(|\psi(0)\rangle = |1\rangle\), then
$$
a(t) = \langle 1|\psi(t)\rangle = \sum_k \exp\left(-\frac{i}{\hbar} E_k t\right) \left|\psi^{(k)}_1\right|^2.
$$
Accordingly, the probability of remaining in the state \(|1\rangle\) at time \(t\) is \(P(t) = |a(t)|^2\).
The plot of this function for the case
\(N = 2000\), \(dE = 10^{-4}\), \(W = 1/3000\), \(\hbar = 1\)
is shown in Fig. \ref{risraspad1}.
Meanwhile, Fig. \ref{risraspad2} illustrates the deviation of \(P(t)\) from the exponential decay law \(\exp(-\Gamma t/\hbar)\).

From Fig. \ref{risraspad1}, it is clearly seen that at \( t \to 0 \), the decay rate is zero, and only over time does \( P(t) \) approach the ``correct'' exponential law.
From Fig. \ref{risraspad2}, it follows that the maximum deviation of \( P(t) \) from the exponential is about 4\% and is reached at \( t \sim 20 \), which agrees well with our estimate for \( T_{min} \). For the parameters we have chosen, \( T_{min} = 10 \). Moreover, this figure clearly shows the oscillations we mentioned earlier.

\begin{figure}
\centerline{\includegraphics[scale=0.7]{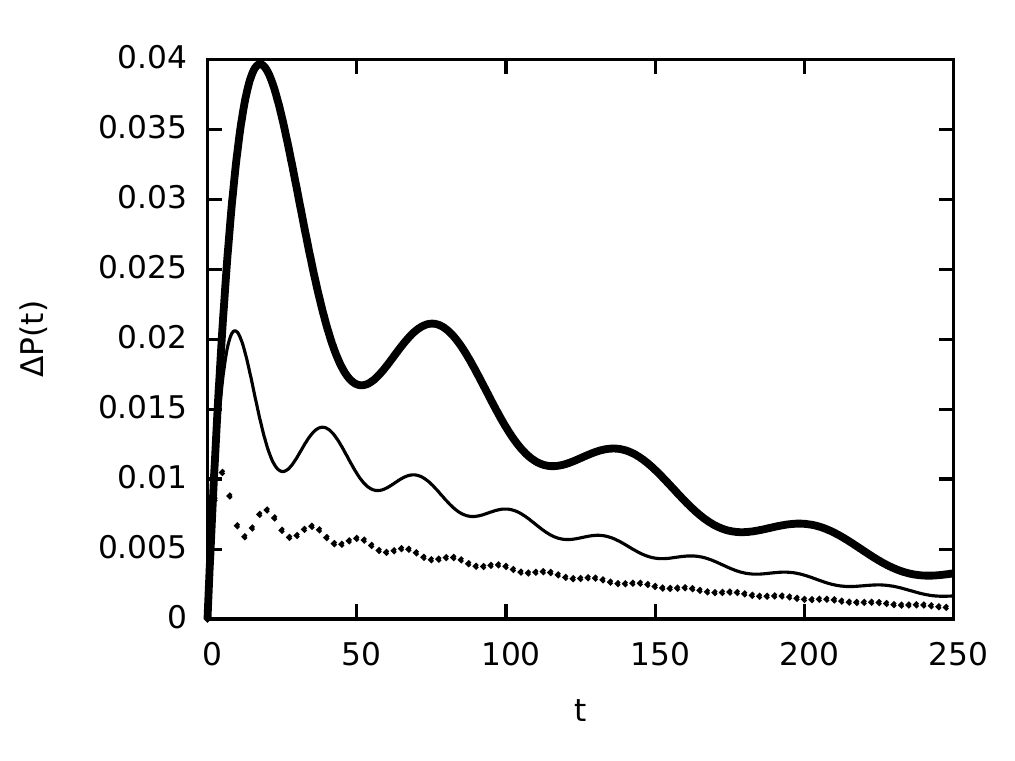}}
\caption{The dependence of the quantity \(\Delta P(t)\) on time \(t\). The bold line corresponds to \(N=2000\), the thin line to \(N=4000\), and the dotted line to \(N=8000\).}
\label{raspadotn}
\end{figure}

If \( N \) is increased while keeping \( R \) fixed, the deviation from the exponential law decreases approximately as \( 1/N \).
This is clearly seen in Fig. \ref{raspadotn}, which shows the plots of the quantity \( \Delta P(t) \equiv P(t) - \exp(-\Gamma t/\hbar) \) for \( N = 2000 \), \( N = 4000 \), and \( N = 8000 \).

\begin{figure}
\centerline{\includegraphics[scale=0.7]{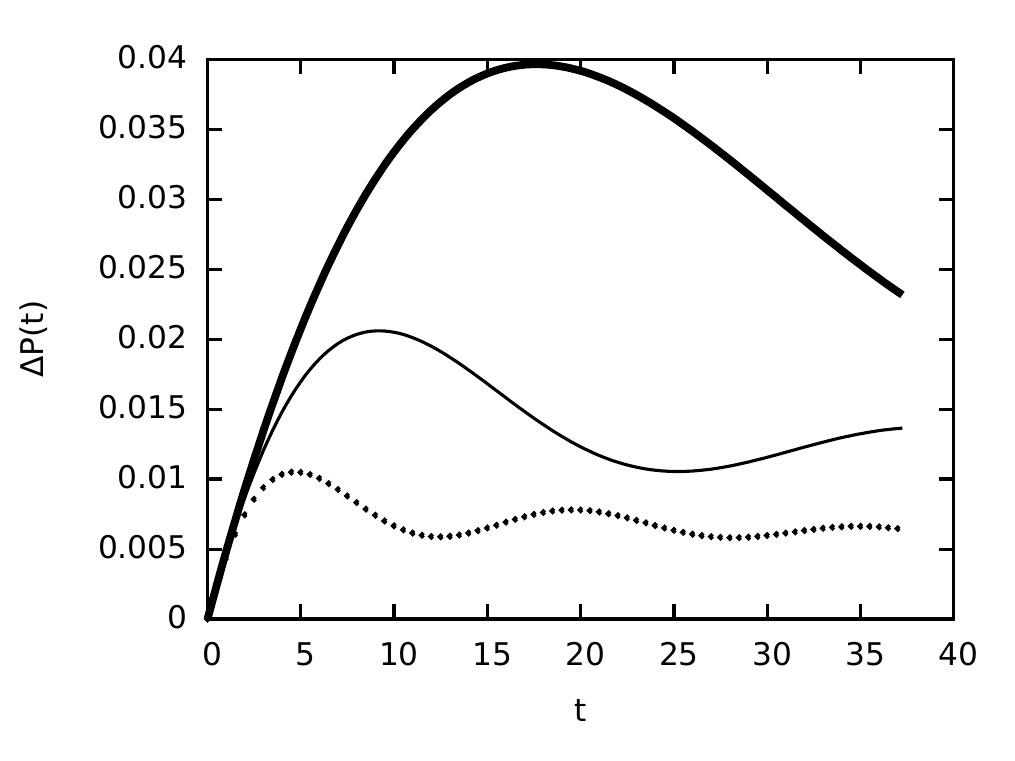}}
\caption{The dependence of the quantity \( \Delta P(t) \) on time at small \( t \). The bold line corresponds to \( N = 2000 \), the thin line to \( N = 4000 \), and the dotted line to \( N = 8000 \).}
\label{raspadotn2}
\end{figure}

At the same time, \( T_{min} \) also decreases as \( 1/N \).
As a result, both the maximum deviation and the size of the region where a noticeable deviation from the exponential law is observed decrease with increasing \( N \).
Figure \ref{raspadotn2} shows the same \( \Delta P(t) \) plots as in Fig. \ref{raspadotn}, but over a shorter time interval.
It is clearly seen that for \( N = 2000 \), the maximum deviation occurs at \( t \sim 20 \) and is about 4\%;
for \( N = 4000 \), the maximum deviation occurs at \( t \sim 10 \) and is about 2\%;
and for \( N = 8000 \), the maximum deviation occurs at \( t \sim 5 \) and is about 1\%.

As for the upper limit on the evolution time, as expected, a partial recovery of the state is observed at \( t \sim T_{max} \).
Due to the perturbation \( H_I \), the spectrum of the Hamiltonian \( H_0 + H_I \) ceases to be equidistant, preventing a complete revival of the initial state at \( t = T_0 = 2\pi\hbar/dE \approx 62831.9 \). However, at \( t = T_0 + 400 \), the initial state is restored by approximately 55\%.
As time progresses further, at multiples of \( T_0 \), the maxima of the function \( P(t) \) become progressively smaller and eventually settle into a kind of ``saturation level'' around 10\%.
It is quite clear that by finding the least common multiple of the quantities \( [10^M \cdot E_k/dE] \) (where the square brackets denote the integer part operation), one can determine the time \( \mathcal{T} \) at which the initial state will be restored with any predetermined accuracy (this accuracy is determined by the integer \( M \)).
However, even for \( N = 2000 \) and \( M = 1 \), this time \( \mathcal{T} \) (which, due to space constraints, we cannot provide here) already lacks any physical meaning and should be regarded as effectively infinite.

\section*{Conclusion}
\mbox{}\vspace{-\baselineskip}

The proposed finite-dimensional model of a metastable level has the following properties:
First, the accuracy of the constructed approximate analytical solution improves with increasing model dimension \( N \) and the dimensionless parameter \( R = \Gamma/dE \), which is the ratio of the linewidth \( \Gamma \) to the spacing between discrete levels \( dE \).
Second, as both \( N \) and \( R \) increase, the accuracy of reproducing the metastable level's line shape (i.e., the Lorentzian curve) improves.
Third, with increasing \( N \) at a fixed \( \Gamma \), the accuracy of reproducing the exponential decay law increases (though, of course, only for \( t \ll T_{max} \)).

We believe that the primary advantage of the proposed model is its simplicity. While in the Fano formalism one has to solve integral equations with singular functions, our model reduces everything to an explicit analytical solution of the spectral problem for a finite-dimensional matrix.

The proposed model can, in particular, be applied to the problem of a quantized fermion field in the presence of a strong electric field. In this problem, the discrete spectrum lies within the energy gap between electron and positron states. However, when the field strength becomes sufficiently large, the discrete level merges into the continuum, giving rise to a resonance.
For weak electric fields, one could use free-field solutions as the continuous spectrum, but the presence of a strong field necessitates the use of nonperturbative solutions --- that is, solutions obtained in the presence of the external field. The conventional approach to renormalization in this problem involves using counterterms corresponding to free fields to renormalize quantities computed with nonperturbative solutions \cite{gyulassy}, \cite{mohr}. However, the self-consistency of this approach raises certain doubts.
Since the finite-dimensional model is, on the one hand, inherently regularized and, on the other hand, some of its results exhibit a smooth limit as \( N \to \infty \), it is possible that this model could help clarify whether the standard renormalization procedure is physically justified or whether it requires certain modifications.


\begin{thebibliography}{9}

\bibitem{fano0}
{\it Fano U.} // Nuovo Cim.  1935. {\bf 12}. P. 154. DOI 10.1007/BF02958288

\bibitem{fano}
{\it Fano U.} // Phys. Rev. 1961. {\bf 124}. P. 1866. DOI 10.1103/PhysRev.124.1866

\bibitem{fonda}
{\it Fonda L.} et al. // Rep. Prog Phys. 1978. {\bf 41}. P. 587. DOI 10.1088/0034-4885/41/4/003

\bibitem{feshbach1}
{\it Feshbach H.,  Porter C.E.,   Weisskopf V.F.} // Phys. Rev.  1954. {\bf 96}.  P. 448. DOI 10.1103/PhysRev.96.448

\bibitem{feshbach2}
{\it Feshbach H.} // 
 Ann. Phys.   1958. {\bf 5}.  P. 357. DOI 10.1016/0003-4916(58)90007-1


\bibitem{kabachnik} {\it Kabachnik N. M., Sazhina I. P. }  //  J. Phys. B.   1976.   {\bf 9}.  P. 1681. DOI 10.1088/0022-3700/9/10/014


\bibitem{limonov1}
{\it Limonov M.F.  et al }  //
Phys. Rev. Lett.   1998.   {\bf  80}.   P. 825. DOI 10.1103/PhysRevLett.80.825


\bibitem{limonov2}
{\it Limonov M.F.  et al } // Phys. Rev. B.   2002.   {\bf  66}.  P. 054509. DOI 10.1103/PhysRevB.66.054509

\bibitem{hopfield} {\it 
 Hopfield J. J., Dean P. J., Thomas D. G.}   // Phys.
Rev.   1967.   {\bf 158}.   P. 748. DOI 10.1103/PhysRev.158.748

\bibitem{cerdeira}
{\it  Cerdeira F., Fjeldly T. A., Cardona M.  }  // Phys. Rev.
B.   1973.   {\bf  8.}   P. 4734. DOI 10.1103/PhysRevB.8.4734

\bibitem{madhavan} {\it Madhavan V.  et al} // Science.   1998.   {\bf 280}.   P. 567. DOI 10.1126/science.280.5363.567

\bibitem{fan}
{\it Fan S.}  // Appl. Phys. Lett.  2002.  {\bf 80}.  P. 908. DOI 10.1063/1.1448174

\bibitem{fan2}
{\it Fan S., Suh W., Joannopoulos J. D. } // J. Opt. Soc. Am. A.   2003.  {\bf 20}. P. 569. DOI 10.1364/JOSAA.20.000569

\bibitem{kong} {\it 
Kong X., Xiao G. }  // J. Opt. Soc. Am. A.  2016.   {\bf 33}.   P. 707. DOI 10.1364/JOSAA.33.000707

\bibitem{galli}  {\it 
Galli M. et al }   // Appl. Phys. Lett.  2009.  {\bf 94}.  P. 071101. DOI 10.1063/1.3080683

\bibitem{smith} {\it 
Smith D. D. et al}   // Phys. Rev. A.  2004. {\bf 69}.  P. 063804. DOI 10.1103/PhysRevA.69.063804

\bibitem{verslegers}
{\it Verslegers L. et al} // Phys. Rev. Lett.  2012.  {\bf 108}. P. 083902. DOI 10.1103/PhysRevLett.108.083902


\bibitem{yang} {\it Yang Y. et al} // Nature Commun.  2014.   {\bf 5}. P. 5753. DOI 10.1038/ncomms6753

\bibitem{peng} 
{\it Peng B. et al} // Nature Commun.  2014. {\bf 5}. 
P. 5082. DOI 10.1038/ncomms6082

\bibitem{gores} {\it Gores J. et al}   // Phys. Rev. B.  2000.   {\bf 62}.   P. 2188. DOI 10.1103/PhysRevB.62.2188

\bibitem{johnson} {\it Johnson A. C. et al}  // Phys. Rev. Lett.  2004.  {\bf 93}.  P. 106803. DOI 10.1103/PhysRevLett.93.106803

\bibitem{grenier1}
{\it Grenier W.} et al. // Z. Physik. 1972. {\bf 257}. P. 62. DOI 10.1007/BF01398198

\bibitem{grenier2}
{\it Grenier W.} et al. // Z. Physik. 1972. {\bf 257}. P. 183. DOI 10.1007/BF01401203

\bibitem{kosta1}
{\it Krasnov A., Sveshnikov K.} // Mod. Phys. Lett. A. 2022. {\bf 37}. P. 2250136. DOI 10.1142/S021773232250136X

\bibitem{kosta2}
{\it Grashin P., Sveshnikov K.} // Int. J. Mod. Phys. A. 2023. {\bf 38}. P. 2350125-1. DOI 10.1142/S0217751X23501257

\bibitem{fermi}
{\it Fermi E.} Lectures on quantum mechanics. University of Chicago Press, 1961.
 
\bibitem{khrust}
{\it Khrustalev O.A., Lunev F.A., Sveshnikov K.A.} et al. Introduction to Quantum Theory. M., 1985.

\bibitem{gyulassy}
{\it Gyulassy M.}  // Nucl. Phys. A. 1975. {\bf 244}. P. 497. DOI 10.1016/0375-9474(75)90554-0

\bibitem{mohr}
{\it Mohr P. J., Plunien G., Soff G. }  // Phys. Rep. 1998. {\bf 293}. P. 227. DOI 10.1016/S0370-1573(97)00046-X

\end{thebibliography}
\end {document}